# Magnetization and magneto-electric effect in La-doped BiFeO$_3$

G. Le Bras, D. Colson, A. Forget, N. Genand-Riondet, R. Tourbot and P. Bonville

CEA, Centre d'Etudes de Saclay, Service de Physique de l'Etat Condensé
91191 Gif-sur-Yvette, France

**Abstract**
We report on magnetisation and magneto-capacitance measurements in the Bi$_{1-x}$La$_x$FeO$_3$ series for $0 \leq x \leq 0.15$. We confirm that doping with La reduces the threshold magnetic field H$_c$ for cancelling the magnetic spiral phase, and we show that H$_c$ decreases as the La content increases up to x=0.15, which is the highest concentration for maintaining the non-centrosymmetric rhombohedral structure of BiFeO$_3$. Measurements of the dielectric constant as a function of magnetic field in the series also show a maximum magneto-capacitance for x=0.15.

**INTRODUCTION**

Recently, the perovskite compound BiFeO$_3$, where ferroelectricity and antiferromagnetism coexist at room temperature, has been the subject of renewed interest[1,2,3,4]. Such multiferroic[5,6] materials can present a strong coupling between the electric and magnetic order parameters, also called the magnetoelectric (ME) effect. Thus they are potential candidates for applications to devices where, for instance, the magnetisation could be switched by the reversal of an electric field[7,8]. However, this is only of interest if some net magnetisation arises in the material. In BiFeO$_3$, the antiferromagnetic (AF) structure is of G-type, but it is not collinear, showing a spiral cycloidal spin arrangement in which the Fe$^{3+}$ moments rotate in a plane, with an incommensurate period $\lambda \cong 620$Å[9,10]. Due to the magnetoelectric coupling, the two Fe$^{3+}$ sublattices are canted, resulting in a weak local net magnetisation, which however averages to zero over a period of the incommensurate modulated spin structure[1]. Another consequence of the cycloidal spin structure is that it inhibits the observation of the linear ME effect[11], i.e. of the linear dependence of the induced electric polarisation on the magnetic field. The ME coupling is however strong enough to allow the reversal of AF domains by switching the electric field, as shown recently[12].

It is therefore of interest to try and destroy the space-modulated spin cycloid. Various methods have been used for this purpose. The application of a magnetic field of 20T has been shown to induce a transition to a homogeneous AF state and to restore a non-zero magnetisation[1,13]. Making BiFeO$_3$ in thin film form also leads to a homogeneous AF spin structure[14,15]. Low level substitutions of rare earth ions (La, Dy, …) for bismuth have an interesting effect: the introduction of rare earth cations in BiFeO$_3$ seems to increase the magnetocrystalline anisotropy[16], thus making the cycloidal spin structure energetically unfavourable. It can then be destroyed by a magnetic field lower than in pure BiFeO$_3$.

La-doped BiFeO$_3$ has already been the subject of numerous investigations[17,18,19,20,21], but the transition from the cycloidal to the homogeneous AF state has not yet been detected in high field magnetisation curves. In this work, we determine the effect of a magnetic field up to 14T on the magnetisation and dielectric constant (or magneto-capacitance) in the Bi$_{1-x}$La$_x$FeO$_3$ series, with x varying between 0 and 0.15, in polycrystalline samples. We show that the critical field for the transition from spiral to AF collinear magnetic structure decreases with increasing La doping, and we observe a correlated behaviour for the magneto-capacitance as a function of magnetic field.

**CRYSTAL STRUCTURE AND EXPERIMENTAL DETAILS**

Ceramic samples of $Bi_{1-x}La_xFeO_3$ (0, 0.05, 0.10, 0.15, 0.20 and 0.25) were synthesized by a conventional solid state reaction using the starting oxides $Bi_2O_3$, $La_2O_3$ and $Fe_2O_3$ (purity ≥99.99%). The mixtures were ground and calcinated at 960°C during 10h. Then the samples were ground and pressed at 6 kbars into pellets 5 mm in diameter and 1.5 mm in height, and sintered at 960°C for 10 hours. All the samples were found single phase, except for a small amount (~2–3%) of $Bi_{25}FeO_{39}$ which was detected for x=0 and x=0.25. For pure $BiFeO_3$, the XRD reflections were indexed in the rhombohedral system (R3c) with lattice parameters a=0.5576(2) nm and c=1.3863(2) nm. Analysis of the XRD patterns (Figure 1) reveals that, when increasing the La content, the lattice symmetry gradually changes from rhombohedral (R3c) to orthorhombic (C222). For instance, the intensity of the (006) and (018) peaks becomes weaker and the splitting of the peaks around 39° and 50° decreases (see resp. insets (a) and (b) in Fig.1), indicating that the rhombohedral distortion is reduced as the La content increases. It has almost completely disappeared near x=0.15. These observations agree with those already published by Zalesskii et al.[22].

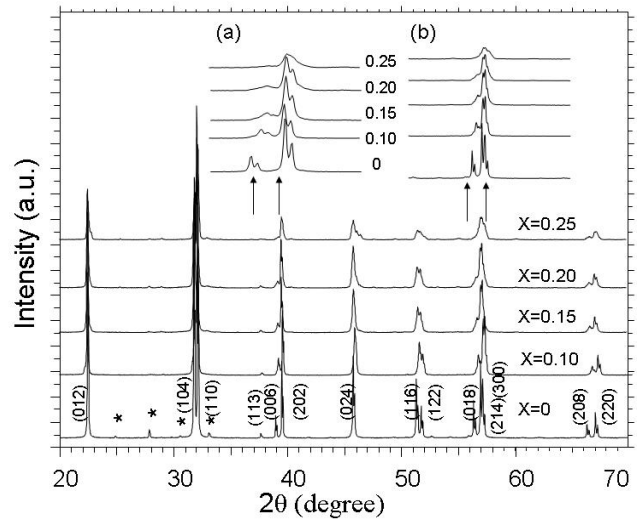

*Figure 1:* Powder X-Ray diffraction pattern in $Bi_{1-x}La_xFeO_3$ (0≤x≤0.25) at 300K using $Cu_{K\alpha}$ radiation. The asterisks correspond to the lines of the $Bi_{25}FeO_{39}$ impurity, especially visible for x=0. Insets: enlarged view near 2θ = 39° (a) and 50° (b) showing the progressive disappearance of the rhombohedral splitting (arrows) with increasing La content.

The isothermal dc magnetization curves were measured at 20 K as a function of the applied magnetic field H up to 14T, using an automated vibrating sample magnetometer (Cryogenic Ltd). It has been checked that the measurements are temperature independent up to 200K.

Dielectric constants were obtained by measuring the capacitance at 10 kHz on the ceramic pellets using a commercial LCR meter (Instek LCR-819). Silver paste (Dupond 4929) painted on both planes was used as electrode. The magnetic field up to 8T was applied perpendicular to the plane of the disk. To avoid ohmic losses due to grain boundaries, the measurements were not performed at room temperature, but at 100K where the capacitance of our samples is nearly T-independent with a value around 20 pF.

## THE SPIRAL STRUCTURE IN $Bi_{1-x}La_xFeO_3$

In pure $BiFeO_3$, the spontaneous electric polarization $\mathbf{P^0}$ below $T_C \approx 1100K$ lies along the cubic $[111]_c$ direction ($[001]_{hex}$), labelled the z–axis in the following, and below $T_N = 673K$, the spiral magnetic structure is such that the $Fe^{3+}$ moments lie in the $(110)_{hex}$ plane, the propagation vector being $\mathbf{q_0} = [110]_{hex}$. Introducing the unit vectors $\mathbf{L}= (\mathbf{M}_1 - \mathbf{M}_2)/2M_0$ and $\mathbf{M} = (\mathbf{M}_1 + \mathbf{M}_2)/2M_0$, where $\mathbf{M}_1$ and $\mathbf{M}_2$ are the $Fe^{3+}$ moments on the two AF sublattices and $M_0$ their common length, and using the formalism of Refs.1,2,3, the magnetoelectric (ME) effect in $BiFeO_3$ results in the presence of energy invariants mixing electric and magnetic variables (see Fig.2), and leading thus to a coupling between them.

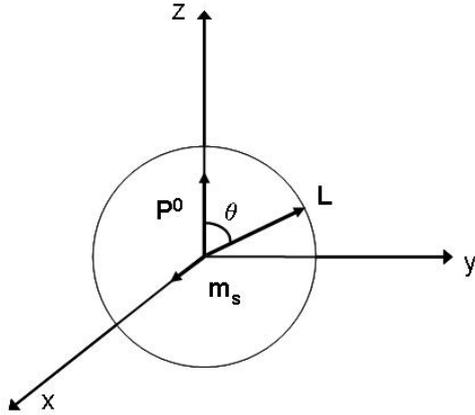

**Figure 2**: *Schematic representation of the magnetic and electric polarisation vectors in $BiFeO_3$. The magnetic spiral is taken to lie in the zOy plane, the angle θ determining the current position of the vector* $\mathbf{L}$. $\mathbf{P^0}$ *is the spontaneous electric polarisation along Oz and* $\mathbf{m_s}$ *the local net moment.*

The ME invariant of the type $g_1 \propto P_z (H_yL_x - H_xL_y)$ leads to a spontaneous magnetic moment:

$$\mathbf{m_s} = - \partial g_1/\partial \mathbf{H} \propto P^0 (L_y, -L_x, 0) \propto P^0 \sin\theta \; \mathbf{i} \qquad (1)$$

along the x-axis (see Fig.2). This spontaneous moment can be obtained in an equivalent manner considering the ME invariant:

$$g_2 \propto P_z (M_yL_x - M_xL_y) = - \mathbf{P^0} \cdot \mathbf{M} \times \mathbf{L} = - \mathbf{P^0} \cdot \mathbf{M}_1 \times \mathbf{M}_2, \qquad (2)$$

emphasizing thereby the similarity with the classical Dzyaloshinsky-Moriya interaction[23,24]. Averaging over a period of the spiral, this moment vanishes. In single crystal $BiFeO_3$, application of a magnetic field along $[001]_c$ induces a transition to a collinear AF structure at $H_c=20T$[13], which is accompanied by the appearance of a weak ferromagnetic moment. In La-doped $BiFeO_3$, in zero field, the spiral structure has been shown to persist up to x=0.10 by nuclear (antiferro)magnetic resonance[22]. From our experiments with $^{57}Fe$ Mössbauer spectroscopy, we could evidence the signature of the spiral structure, i.e. a characteristic spectral asymmetry[25], only up to x=0.05. For higher dopings, the line broadenings due to cation disorder seem to preclude the observation of the spiral induced asymmetry. Thus the spiral magnetic structure is likely to be present as long as the crystal structure remains rhombohedral, i.e. up to x=0.15.

## RESULTS
### 1) Magnetization measurements

The isothermal magnetisation curves for various La doping contents $0 \leq x \leq 0.15$ in $Bi_{1-x}La_xFeO_3$ polycrystalline samples are shown in Fig. 3a. At low and intermediate fields, for pure $BiFeO_3$, the m(H) curve exhibits a linear field dependence, as expected for an antiferromagnetic spin structure. However, a departure from the linear behaviour appears for H~6T, and for H≥ 6T the field dependence is noticeably non linear. With increasing La content (up to x=0.15), the low field magnetization is still linear, but the threshold field for deviation from the linear behaviour decreases. For x=0.10, 0.125 and 0.15, a linear dependence is the recovered at high field, with a somewhat higher slope. We interpret this behaviour as due to the crossover from the spiral magnetic structure to a collinear one when the field is increased. This crossover takes place for decreasing fields as the La doping increases.

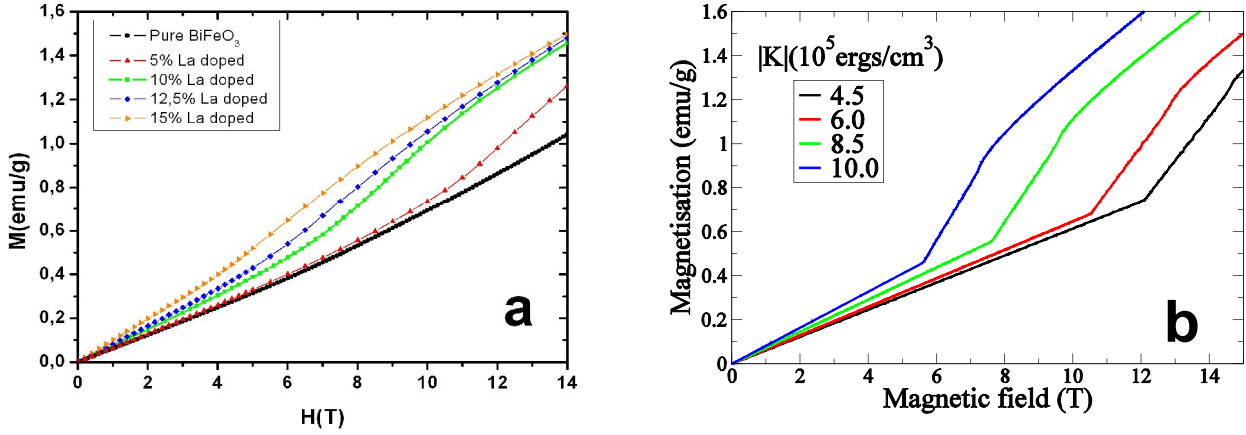

*Figure 3: a) Magnetisation curves at 20K in $Bi_{1-x}La_xFeO_3$ ceramic samples for $0 \leq x \leq 0.15$. The lines are guide for the eyes; b) Calculated m(H) curves in a polycrystalline sample for different values of the anisotropy energy density K.*

The width of the transition (a few T) is due to the dependence of the critical field upon its orientation with respect to the crystal axes, leading to a spread of $H_c$ values in our ceramic samples.

In order to interpret more quantitatively the magnetisation curves, we computed the orientational dependence of the critical field $H_c$ using the theory developed in Refs.1, 2, 3. For this purpose, one has to assume some spin configuration in the high field collinear phase, which is likely to depend on the type of crystalline (small) anisotropy in $BiFeO_3$. Recently, the density functional theory has been applied to $BiFeO_3$[26] and the anisotropy was found to be of planar type, i.e. the easy plane is the $(001)_{hex}$ plane. Using an approach similar to that of Ref.27, the same result is obtained and, assuming the single ion anisotropy density is written $E_a = K \sin^2\theta$, a numerical estimate yields $K \approx -2 \times 10^6$ ergs/cm$^3$ (see Ref.28). Therefore, we make the reasonable assumption that, in the collinear AF phase, the **L** vector remains in the basal xOy easy plane and is oriented perpendicular to the field. Then, for a given orientation $(\psi,\gamma)$ of the applied field, computation of the critical field is straightforward and is given in detail in the Appendix. The parameters of the problem are the transverse AF susceptibility $\chi_\perp$, the anisotropy density K, the gain in exchange energy density of the spiral phase $E_s = Aq_0^2$, where A is the exchange stiffness, and the ME-induced spontaneous magnetisation $m_s$. The critical field $H_c(\psi,\gamma)$ depends on $m_s/\chi_\perp$ and on $(Aq_0^2 - |K|/2)/\chi_\perp$. For a given applied magnetic field, we compute the powder magnetisation using:

$$M(H) = \frac{1}{4\pi} \int_0^{2\pi} d\gamma \int_0^{\pi} d\psi \, \sin\psi \, F(H,\psi,\gamma), \qquad (3)$$

with, using equations (A2) and (A4):

$$F(H,\psi,\gamma) = \tfrac{1}{2} \chi_\perp H (\sin^2\psi \cos^2\gamma + 1) \quad \text{for } H < H_c(\psi, \gamma), \qquad (4)$$
$$F(H,\psi,\gamma) = \chi_\perp H + m_s \sin\psi \quad \text{for } H > H_c(\psi, \gamma).$$

The asymptotic forms for m(H) after powder averaging are:

$$m(H) = 2/3\, \chi_\perp H \text{ for } H < \min\{H_c(\psi,\gamma)\}, \qquad (5)$$

i.e. when all crystallites are in the spiral phase, with no ferromagnetic component as expected; and

$$m(H) = \chi_\perp H + \pi/4\, m_s \text{ for } H > \max\{H_c(\psi,\gamma)\}, \qquad (6)$$

i.e. when all crystallites are in the collinear phase. These expressions account for the increase of the *m(H)* slope observed for the samples with x ≥ 0.10 between the low and high field regions; the model also predicts a weak ferromagnetic moment above the critical field, which is observed for the sample with x=0.15. The dependence of $H_c$ on the anisotropy density K can be examined for instance in a simple case, i.e. when the field is applied along the x-axis. Then (see (A6)):

$$H_c = \tfrac{1}{4}\, (m_s/\chi_\perp)\, [4\, \chi_\perp\, (Aq_0^2 - |K|/2)\, /\, m_s^2 - 1].$$

It appears from this expression that $H_c$ is a decreasing function of *|K|*, as can be expected from simple energetic considerations. This trend holds for any orientation of the applied field. Experimentally, the low field slope $2/3\chi_\perp$ is seen to increase slightly with La doping, yielding $\chi_\perp$ values between $7.5\times10^{-5}$ for pure $BiFeO_3$ and $10.0\times10^{-5}$ for x=0.125. While we have no explanation as to the reason for this increase, we have included it in our simulations. Figure 3b represents the m(H) curves computed for different values of |K|, the energy gain for the spiral state being [29]: $E_s = Aq_0^2 = 6.5\times10^5$ ergs/cm$^3$. We find that |K| values ranging from $4.5\times10^5$ to $10^6$ ergs/cm$^3$ and a weak ferromagnetic magnetisation $m_s = 1.2$ emu/cm$^3$ = 0.15 emu/g reproduce qualitatively the observed features of the m(H) curves with increasing La doping. Although they do not match precisely the experimental data, these calculations put on a more solid ground the assumption of an increasing anisotropy density with increasing La content in the rhombohedral $Bi_{1-x}La_xFeO_3$ compounds.

### 2) Magnetodielectric measurements

In Figure 4a is reported the magnetic field dependence of the normalized dielectric constant ε(H)/ε(H=0) of samples with different La contents at 100K.

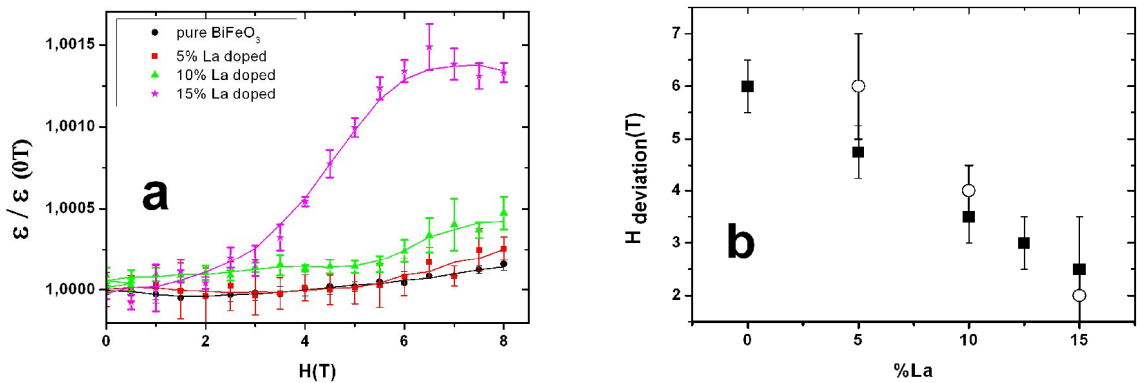

*Figure 4*: In the $Bi_{1-x}La_xFeO_3$ series: *a)* Magnetic field dependence of the normalized dielectric constant at 100K; *b)* Threshold fields for deviation of m(H) from a linear law (full squares) and for deviation of ε(H) from a constant (open circles).

In pure BiFeO$_3$, as well as for x=0.05, there is no clearcut field dependence of the dielectric constant up to 8T. For x=0.10, a slight increase of ε above 5T is observed, and for x=0.15 a clear upturn occurs above 3T, culminating in a 0.14% relative enhancement around 8T. We associate the stronger field dependence of ε as x increases up to 0.15 with the destruction of the spiral spin structure at lower fields, as shown by the preceding magnetisation data. Figure 4b shows the comparison of the experimentally determined threshold fields for deviation of the m(H) curves from a linear law and for deviation of the ε(H) curves from a constant value, as a function of La content. There is a clear correlation, which is a strong indication that the two phenomena are linked. In the following, we interpret the behaviour of ε(H) in a way similar to the cancellation of the linear magnetoelectric effect by the spiral magnetic structure and its recovery in the presence of a strong magnetic field.

In the presence of magnetic and electric fields, the *k* component of the induced polarisation in a multiferroic material reads, up to the second order in the fields (and using the Einstein rule of summation over repeated indices):

$$P_k(E,H;T) = -\frac{\partial g}{\partial E_k} = {}^sP_k + \varepsilon_0 \varepsilon_{ik} E_i + \alpha_{ki} H_i + \frac{1}{2}\beta_{kij} H_i H_j + \gamma_{ijk} H_i E_j \quad (8)$$

where $^sP$ is the spontaneous electric polarisation, $\varepsilon_0$ the free space permittivity. The second rank tensors $\varepsilon_{ik}$ and $\alpha_{ik}$ are resp. the relative permittivity and the linear magnetoelectric coupling, and the third rank tensors $\beta_{ijk}$ and $\gamma_{ijk}$ are associated with higher order magnetoelectric interactions[30,31,32]. In our ac measurement of the dielectric constant (or of the relative permittivity), the magnetic field **H** is static and the electric field **E** is oscillating. Therefore, in order to obtain the ac induced polarisation, we must keep only the terms proportional to **E** in the expression for **P**. Then, the H-dependent renormalised permittivity tensor writes:

$$\varepsilon_0 \Sigma_{ij}(\mathbf{H}) = \varepsilon_0 \varepsilon_{ij} + \tfrac{1}{2} \gamma_{kij} H_k, \quad (9)$$

i.e. it depends linearly on the magnetoelectric tensor **γ**. The **α** and **γ** tensors can be derived from energy invariants containing resp. terms of the type $E_i H_k L_j$ and $H_i E_j E_k L_m$, i.e. they depend linearly on the AF vector **L**. In the spiral AF phase of BiFeO$_3$, <**L**> = **0** over a period of the spiral, implying <**α**> = <**γ**> = 0. Therefore, there is no linear H-dependence of the dielectric constant, as experimentally observed. On increasing the La content, the spiral structure is progressively destroyed in lower fields, leading to a measurable H-dependence of the dielectric constant. Averaging over the orientations in our ceramic sample should not alter the main conclusions derived from expression (9), which provides thus a qualitative explanation of the observed behaviour of ε(H).

### DISCUSSION

We considered here only a negative sign for the anisotropy constant K, i.e. an easy plane perpendicular to the electric polarisation, for the Fe$^{3+}$ moments. This is in line with recent calculations[26,28] and the numerical estimate from *ab initio* calculations[28] |K| ≈ 2x10$^6$ ergs/cm$^3$ is reasonably close to the values we used for our simulations. Our values are also similar to that derived from electron spin resonance experiments in Ref.29, where however K is taken to be positive. We remark that the sign of K is irrelevant as to the stability of the spiral structure as long as the condition: $-Aq_0^2 + |K|/2 < 0$ is fulfilled, i.e. the spiral structure is stable if $|K| < 2Aq_0^2$.

It is worth noting that, besides the present spin configuration in the high field collinear phase, we tried other reasonable assumptions, but none of them yielded a solution for the critical field for all field orientations in the crystal axes. This spin configuration, for which **L** is oriented perpendicular to **H**, implies that the powder susceptibility in the high field phase is exactly $\chi_\perp$ (see expression (6)). Comparison with experimental results shows that this value is over-estimated, which could be due to the crudeness of our assumption. Neutron scattering with a magnetic field could help determine the high field magnetic structure if large enough crystals of the doped materials were available.

The magnetic and dielectric properties of the x=0.15 compound have recently been studied in Ref.33. In their sample, the authors find a crystal structure very close to orthorhombic, without rhombohedral distortion, in contradiction with our findings. We think this result comes from the fact that x=0.15 is a borderline composition, and that small deviations from this nominal value can push the material on one or either side of the rhombohedral to orthorhombic transition.

What are the possible mechanisms for the increase in anisotropy with La doping? Since the presence of orbital degrees of freedom, like in $Co^{3+}$, $Fe^{2+}$ or the rare earths, leads effectively to large K values, one could think of the presence of $Fe^{2+}$ (associated to oxygen vacancies) or to an admixture of *4f* components into the Fe wave-functions, both effects arising from the presence of La. On the one hand, our $^{57}$Fe Mössbauer measurements in the doped compounds show that the $Fe^{2+}$ content (if any) is very small, lower than 1%. On the other hand, a *d-f* hybridisation at the Fe site could be conceivable in a metallic or strongly covalent material, but its presence in an ionic compound like $BiFeO_3$ seems rather unlikely. Thus the exact mechanism for the enhancement of K due to La-doping remains to be found. We note that another effect has been found, in zero field, when doping $BiFeO_3$ with $Mn$[34]: an increase of the spatial period λ of the spiral spin modulation, together with a "bunching" of the Fe moments towards the $[001]_{hex}$ axis.

**CONCLUSION**

We have performed magnetic and dielectric measurements as a function of magnetic field in La-doped $BiFeO_3$ polycrystalline samples. We confirm that La substitution reduces the transition field $H_c$ from the spatially modulated to the collinear magnetic state and we find a good correlation between the $H_c$ values obtained by magnetic and dielectric measurements as a function of La doping. In our ceramic samples, the orientational dependence of the critical field somewhat broadens the transition, but we show that the transition is globally shifted to lower fields with increased La doping. Calculations based on the free energy model of Refs.1,3 allow to interpret this behaviour in terms of increasing (negative) anisotropy density energy K. In the sample $Bi_{0.85}La_{0.15}FeO_3$, where the rhombohedral structure is still present, we obtain the minimum field (7-8T) for complete transition to the collinear AF phase and the maximum positive magneto-capacitance effect at 8T.

**Acknowledgements**
This work was supported by NTOS-1-45147 "FEMMES".

**APPENDIX**

Our calculation follows the formalism developed in Ref.1 and 3. It uses a simplified model to obtain the critical field for transition from the spiral magnetic phase to the collinear AF phase for an arbitrary orientation (ψ,γ) of the applied field **H,** in the frame defined in Fig.2. Whereas it is known that the cycloidal spiral is harmonic (i.e. sine-wave) at room temperature and in zero field, some anharmonicity sets in on lowering the temperature and/or on increasing the magnetic field. We shall neglect here these departures from the sine-wave cycloid. Apart from the exchange, anisotropy and Zeeman terms in the energy, we write the ME coupling as an effective Zeeman energy with a field $\mathbf{H_{ME}} = \beta \mathbf{L} \times \mathbf{P^0}$, which adds to the applied field. In the spiral phase, the exchange energy density gain over the collinear phase is $E_s = Aq_0^2$. The anisotropy energy density is $E_a = K \sin^2\theta$, with K < 0 as recently determined[26,28]. Averaging over a period, in the harmonic approximation for the cycloid, yields $\langle E_a \rangle = \frac{1}{2} K$. In the collinear phase, our assumption is that **L** lies in the basal xOy plane ($\theta = \pi/2$) and is perpendicular to **H**; hence **L** = (-sin$\gamma$, cos$\gamma$, 0) and $E_a = -|K|$. The Zeeman energy is written: $E_Z = -1/2 \chi_\perp H_\perp^2$, where $H_\perp$ is the component of the total field **H**+**H**$_{ME}$ perpendicular to **L** (we neglect $\chi_{//}$ since T << $T_N$). Introducing the amplitude of the ME-induced magnetisation along Ox in the spiral phase $m_s$, then: $m_s = \chi_\perp H_{ME} = \chi_\perp \beta P^0$, whence the ME coupling parameter $\beta$ is worth: $\beta P^0 = m_s/\chi_\perp$. After a straightforward calculation, one obtains the energy in the spiral phase (after averaging over a period):

$$E_{SP} = - Aq_0^2 - 1/2\, |K| - 1/4\, \chi_\perp H^2 (1 + \sin^2\psi \cos^2\gamma) - 1/4\chi_\perp (m_s/\chi_\perp)^2 \quad (A1)$$

and the magnetisation:

$$M_{SP} = 1/2\, \chi_\perp H\, (1 + \sin^2\psi \cos^2\gamma); \quad (A2)$$

in the collinear AF phase, the energy is:

$$E_{AF} = -|K| - \tfrac{1}{2}\chi_\perp H^2 - \tfrac{1}{2}\chi_\perp (m_s/\chi_\perp)^2 - m_s H \sin\psi \quad (A3)$$

with the magnetisation:

$$M_{AF} = \chi_\perp H + m_s \sin\psi. \quad (A4)$$

With reasonable values of the parameters, $E_{SP}$ is lower than $E_{AS}$ at low fields; hence, above the critical field $H_c$, the collinear phase becomes the ground state. The second order equation that yields the critical field is:

$$(1 - \sin^2\psi \cos^2\gamma)\, H_c^2 + 4(m_s/\chi_\perp) H_c \sin\psi + (m_s/\chi_\perp)^2 - 4/\chi_\perp (Aq_0^2 - |K|/2) = 0. \quad (A5)$$

If $\psi = \pi/2$ and $\gamma = 0$ or $\pi$ (i.e. **H** is along the x-axis), the equation is first order and the critical field writes:

$$H_c(\pi/2, 0 \text{ or } \pi) = \tfrac{1}{4} (m_s/\chi_\perp) [4\, \chi_\perp (Aq_0^2 - |K|/2) / m_s^2 - 1]. \quad (A6)$$

For an orientation of **H** outside the x-axis, the solution for $H_c$ is:

$$H_c(\psi,\theta) = 2\, (m_s/\chi_\perp) (\sqrt{N} - 1) \sin\psi / (1 - \sin^2\psi \cos^2\gamma) \quad (A7)$$

where $N = 1 - [1 - 4\, \chi_\perp (Aq_0^2 - |K|/2)/m_s^2] (1 - \sin^2\psi \cos^2\gamma) / (4 \sin^2\psi) \quad (A8)$

**References**

[1] A. Kadomtseva, A. Zvezdin, Y. Popov, A. Pyatakov, and G. Vorob'ev, JETP Lett. **79**, 571 (2004) and references therein.
[2] A. K. Zvezdin, A. P. Pyatakov, Conferences and symposia Physics-Uspekhi **47**, 416 (2004)
[3] A. M. Kadomtseva, Yu F. Popov, A. P. Pyatakov, G. P. Vorob'ev, A. K. Zvezdin and D. Viehland, Phase Trans. **79,** 1019 (2006)
[4] C. Ederer, C. J. Fennie, J. Phys.: Condens. Matter **20**, 434219 (2008)
[5] M. Fiebig J. Phys. D: Appl. Phys. **38,** R123 (2005)
[6] G. A. Smolenskii and I. Chupis, Sov. Phys. Usp. **25**, 475 (1982)
[7] W. Eerenstein, N. D. Mathur, and J. F. Scott, Nature (London) **442**, 759 (2006). [MEDLINE]

[8] Z.V. Gabbasova, M. D. Kuz'min, A. K Zvezdin, I. S. Dubenko, V. A. Murashov, D. N. Rakov, Phys. Lett. A **158**, 491 (1991)

[9] I. Sosnowska, T. Peterlin-Neumaier, and E. Steichele, J. Phys. C **15**, 4835 (1982).

[10] I. Sosnowska, M. Loewenhaupt, W. I. F. David, and R. Ibberson, Physica B **180&181**, 117 (1992).

[11] Yu. F. Popov, A. K. Zvezdin, G. P. Vorob'ev, A. M. Kadomtseva, V. A. Murashev, and D. N. Rakov, JETP Lett. **57**, 69 (1993)

[12] D. Lebeugle, D. Colson, A. Forget, M. Viret, A. M. Bataille and A. Gukasov, Phys. Rev. Lett. **100**, 227602 (2008)

[13] Yu. F. Popov, A.M. Kadomtseva, A.K. Zvezdin, G.P. Vorob'ev, and A.P. Pyatakov, in *Magnetoelectronic Phenomena in Crystals*, ed. M. Fiebig (Kluwer Academic Publishers, Dordrecht, 2004).

[14] Y. H. Chu *et al*, Nature Materials **7**, 478 (2008).

[15] F. Bai, J. Wang, M. Wuttig, J. F. Li, N. Wang, A. Pyatakov, A. K. Zvezdin, L. E. Cross, D. Viehland, Appl. Phy. Lett. **86,** 182905 (2005)

[16] G. P. Vorob'ev, A. K. Zvezdin, , A. M. Kadomtseva, Yu. F. Popov, V. A. Murashov, and Yu. P. Chernenkov, Phys. Solid State **37**, 1329 (1995)

[17] I. Sosnowska, R. Przenioslo, P. Fischer, V. A. Murashov, J. Magn. Magn. Materials **160**, 384 (1996).

[18] J. R. Sahu, C. N. R. Rao, Solid State Sciences **9**, 950 (2007)

[19] Y. H. Lin, Q. Jiang, Y. Wang, C-W. Nan, L. Chen, J. Yu, Appl. Phys. Lett. **90**, 172507 (2007).

[20] Z. X. Cheng, A. H. Li, X. L. Wang, S. X. Dou, K. Ozawa, H. Kimura, S. J. Zhang and T. R. Shrout, J. of Appl. Phys. **103** 07E507 (2008)

[21] S. T. Zhang, L. Pang, Y. Zhang, M. Lu and Y. Chen, J. of Appl. Phys. **100** 114108 (2006)

[22] A. V. Zalesskii, A. A. Frolov, T. A. Khimich, and A. A. Bush, Phys. Solid State **45**, 141 (2003).

[23] I. E. Dzyaloshinskii, Sov. Phys. JETP **5**, 1259 (1957)

[24] T. Moriya, Phys. Rev. **120**, 91 (1960)

[25] D. Lebeugle, D. Colson, A. Forget, M. Viret, P. Bonville, J. F. Marucco, and S. Fusil, Phys. Rev. B **76**, 024116 (2007).

[26] C. Ederer, N. A. Spaldin, Phys. Rev. B **71**, 060401(R) (2005)

[27] C. J. Fennie, Phys. Rev. Lett. **100**, 167203 (2008)

[28] C. J. Fennie, private communication

[29] B. Ruette, S. Zvyagin, A. P. Pyatakov, A. Bush, J. F. Li, V. I. Belotelov, A. K. Zvezdin, and D. Viehland, Phys. Rev. B **69**, 064114 (2004)

[30] H. Schmid, in *Introduction to Complex Medium for Optics and Electromagnetics*, eds. W. S. Weiglhofer and A. Lakhatia, SPIE Press, USA, 167-195 (2003)

[31] E. Ascher, Phil Mag. **17**, 149 (1968)

[32] H. Grimmer, Ferroelectrics **161**, 181 (1994)

[33] G. L. Yuan, K. Z. Baba-Kishi, J. –M. Liu, S. W. Or, Y. P. Wang and Z. G. Liu, J. Am. Ceram. Soc. **89**, 3136 (2006)

[34] I. Sosnowska, W. Schäfer, W. Kockelmann, K. H. Andersen and I. O. Troyanchuk, Applied Physics A **74** [Suppl.], S1040-S1042 (2002)